# GUI Database for the Equipment Store of the Department of Geomatic Engineering, KNUST

J. A. Quaye-Ballard[1], R. An[2], A. B. Agyemang[3], N. Y. Oppong-Quayson[4] and J. E. N. Ablade[5]

[1]PhD Candidate, College of Earth Science and Engineering, Hohai University, Nanjing, China.
[2]Professor, Dean, College of Earth Science and Engineering, Hohai University, Nanjing, China.
[1,3]Lecturer, Department of Geomatic Engineering, College of Engineering, Kwame Nkrumah University of Science and Technology (KNUST), Kumasi, Ghana
[4,5]Research Assistant, Department of Geomatic Engineering, College of Engineering, KNUST, Kumasi, Ghana

*Abstract*— The geospatial analyst is required to apply art, science, and technology to measure relative positions of natural and man-made features above or beneath the earth's surface, and to present this information either graphically or numerically. The reference positions for these measurements need to be well archived and managed to effectively sustain the activities in the spatial industry. The research herein described highlights the need for an information system for the Land Surveyor's Equipment Store. Such a system is a database management system with a user-friendly graphical interface. This paper describes one such system that has been developed for the Equipment Store of the Department of Geomatic Engineering, Kwame Nkrumah University of Science and Technology (KNUST), Ghana. The system facilitates efficient management and location of instruments, as well as easy location of beacons together with their attribute information; it provides multimedia information about instruments in an Equipment Store. Digital camera was used capture the pictorial descriptions of the beacons. A Geographic Information System (GIS) software was employed to visualize the spatial location of beacons and to publish the various layers for the Graphical User Interface (GUI). The aesthetics of the interface was developed with user interface design tools and coded by programming. The developed Suite, powered by a reliable and fully scalable database, provides an efficient way of booking and analyzing transactions in an Equipment Store.

## I. INTRODUCTION

Scientists around the world are addressing the need to increase access to research data [1]. Software Applications such as the Dataverse Networ, which is an open-source Application for publishing, referencing, extracting and analyzing research data, have been developed to solve the problems of data sharing by building technologies that enable institutions to reduce the burden for researchers and data publishers, and incentivize them to share their data [2]. One of the technologies that most people have become accustomed to hearing about, at work or on the internet is Database. A database is a structured collection of records or data that is stored in a computer system [3]. Accessibility and storage of large amount of data is important for a designing a database system [4-7]. [4] highlights on the requirements for efficient Database Management System (DBMS). In recent years, the object-oriented paradigm has been applied in areas such as engineering, spatial data management, telecommunications, and various scientific domains. This is a programming paradigm using object-oriented data structures consisting of data fields and their links with various methods to design applications and computer programs [7, 8]. Programming techniques may include features such as data abstraction, encapsulation, measuring, modularity, 3D visualization, polymorphism and inheritance [3]. The conglomeration of Object-Oriented Programming (OOP) and database technology has led to this new kind of database. These databases bring the database world and the application-programming world closer together, in particular by ensuring that the database uses the same type of system as the application program. A database system needs be managed with a Graphical User Interface (GUI), which is a Human-Computer Interface (HCI), that is, a way for humans to interact with computers. The GUI employs windows, icons and menus which can be manipulated by use of a mouse and often to a limited extent by use of a keyboard as well. Thus, a GUI uses a combination of technologies and devices to provide a platform for the tasks of collection and producing information [9]. To support the idea of this research, many interesting databases associated with GIS have been developed in many parts of the world to support decision makers in their investigations [e.g. 10-13].

A series of elements conforming to a visual language have evolved to represent information stored in computers. This makes it easier for people with few computer skills to work with and use computer software. The most common combination of such elements in GUIs is the Window Icon Menu Pointing (WIMP) device paradigm, especially in personal computers [9]. A major advantage of GUIs lies in their capacity to render computer operation more intuitive and, thereby, easier to learn and use. For example, it is much easier for a new user to move a file from one directory to another by dragging its icon with the mouse than by having to remember and type practically mysterious commands to accomplish the same task. The user should feel in control of the computer, and not the other way around. This is normally achieved in software applications that embody the qualities of responsiveness, permissiveness and consistency.

Sustaining the activities of the Geoinformation industry, for example, calls for proper management of coordinates to which survey measurements are referred as well as effectual keeping of records. KNUST is a renowned university with adequate resources for running a DBMS at various departmental levels and, in recognition of the urgent need for crossing over from analogue to relevant digital methodologies this work was undertaken to produce a prototypical DBMS for the Department of Geomatic Engineering equipment store. In the latter part of the 1980s significant losses were incurred during a fire outbreak in the block that used to house the departments equipment. The effects of those losses are still felt and to support this need for a computerized record-keeping routine with regular back-ups. Furthermore the extremely large volume of equipment and other stuff makes data retrieval and Instrument location quite difficult in the store. In addition, locating beacons on campus is also a major problem. This calls for development of a database (of all beacons on KNUST campus), and to have such database integrated in a Geographic Information System (GIS) to facilitate identification and locating of beacons. The GIS would integrate modules for capturing, storing, analyzing, managing, and presenting data linked to location; a merging of cartography, statistical analysis, and database technology [14]. With regards to meeting the above-mentioned objectives, a database of all equipment in the department's Equipment Store has been developed with a graphical user interface of all equipment shelf locations. It is a fully scalable GUI database of beacons on KNUST campus. Software has also been developed for record-keeping (of the use of all equipment in the Store) in a digital environment; certain survey computational procedures have also been developed in the application to use beacon coordinate information seamlessly.

## II. MATERIALS AND METHODS

A questionnaire was issued to help study how records are kept at the Equipment Store; how beacons can be easily identified in the field; how surveyors obtain beacon information for their jobs; and how GIS could be used to improve procedures for lending of equipment and to better manage information on beacons. The data used for the research included the beacons' Coordinate data, which include their Descriptions, Northings, Eastings, Elevations and Revision Dates; a video coverage of the Geomatic Equipment Store; pictures and videos of various sections of the Store. The programming languages and Application software employed were Visual Basic (vb.net) for the development of graphical user interface Applications; C-Sharp (C#) for encompassing imperative, declarative, functional, generic, object-oriented and component-oriented programming disciplines. Extensible Application MarkUp Language (XAML) was used for creating user interfaces in WPF and for serializing .NET object instances into a human readable format. Adobe Photoshop was also used for its extensive amount of graphics editing features. Other Applications employed include the .NET framework for building the software application; Windows Presentation Foundation (WPF), a graphical display system designed for the .NET, framework; the Entity Framework for accessing data and working with the results in addition to DataReaders and DataSets; and Environmental Systems Research Institute (ESRI) ArcObjects Library to support the GIS application on the desktop. The logical structure of the methods used for the design of the application is depicted in Fig. 1.

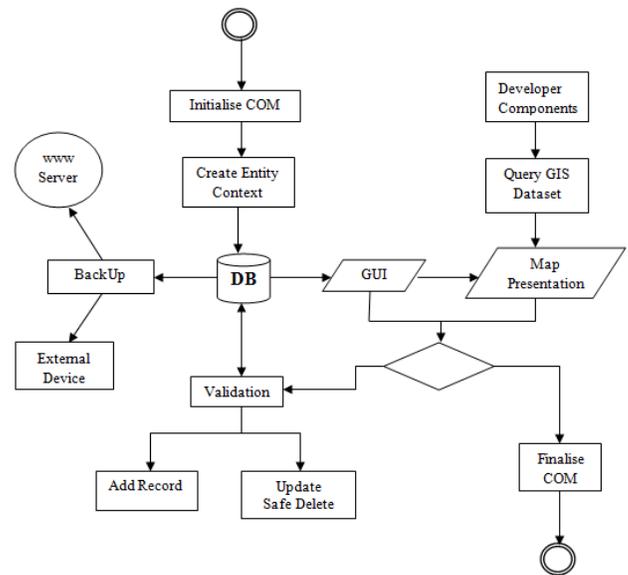

Fig. 1: Logical structure of methods

When an application starts, the Component Object Module (COM) will be initialized. An entity data context is created for the database and its objects are loaded onto the GUI by means of data binding. The GUI running on .NET Framework connects to a GIS dataset via COM Interop. Whenever a record needs to be updated, it undergoes validation. Backing-up the database can be done onto a web server or an external device.

### A. Procedure

The whole suite comprises four applications, namely *BeaconBase*, *LendingBase*, *InstruVisio* and *Survcom*. The BeaconBase application has a database of all the survey beacons on campus with their coordinates in the Ghana National Grid System. The LendingBase application is used for recording lending transactions that occur in the Equipment Store, with details of the borrower and all instruments borrowed. The layout controls used for both BeaconBase and LendingBase applications include the DockPanel, Grid, StackPanel and the Tab Control. The root layout element used is the Grid. Two DockPanels are docked at the Top of the Window and Stretched Horizontally across the screen. The InstruVisio is a multimedia and interactive interface for the instruments stored at the Equipment Store. Multimedia files were encoded in Expression Encoder software for InstruVisio interface. Other controls used were defined in XAML code for InstruVisio interface.

The SurvCom is an application created on the Multiple Document Interface (MDI) architecture of the Windows Form platform. Most of the procedures for developing the application were as in the procedures for the other applications mentioned earlier. MDI applications facilitate the display of multiple documents or forms at the same time, with each document displayed in its own window. MDI applications often have a Window menu item with submenus for switching between windows or documents.

The SQL server compact 3.5 database was employed in designing the LendStore Database (a database for lending), which has ten fields - PersonID which is the primary key field (type integer), Date (type datetime), PersonName (type nvarchar), PersonDepartment (type nvarchar), PersonPhone (type nvarchar), IsReturned (type Boolean), ReturnDate (type datetime), Remarks (type nvarchar), and TotalInstru (type integer); LendingDetails Table, which has five fields: Id which is the primary key field (type integer), InstrumentName (type nvarchar), Quantity (type integer), PersonID which is a foreign key field (type integer) and Serial (type nvarchar); and the Instrument Table, which has six fields: ID which is the primary key field (type integer), InstrumentName (type nvarchar), Unused (type nvarchar), Used (type nvarchar), Remaining (type nvarchar) and Description (type nvarchar); and the BeaconDB Database (a database for the beacons), which has one table with nine fields: BeaconID (type integer), BeaconName (type nvarchar), Northing (type nvarchar), Easting (type nvarchar), Elevation (type nvarchar), Description (type nvarchar), Photo (type nvarchar), RevisionSurveyor (type nvarchar) and Date (type datetime).

The Microsoft Expression Blend and the Visual Studio Integrated Development Environment (IDE) were the GUI design and coding software used respectively. A new project was created for each of the applications under the suite. Expression Blend was employed in the design of the WPF windows. The query language used is the Language-Integrated Query (LINQ). LINQ-To-Entities and Entity Structure Query Language (SQL) are a groundbreaking innovation in Visual Studio and the .NET Framework version 3.5 that bridges the gap between the world of objects and the world of data. In a LINQ query, one always works with objects, and that suited the object-oriented approach used in this work.

*B. Ethical Considerations*

Security issues were considered when developing the software suite. The data in the database was protected against unauthorized disclosure, alteration and destruction. The user is allowed to do only what is expected to be done to avoid inconsistencies in the database which could produce errors in the software. Data integrity or accuracy/correctness of data in the software was also considered. Though this attribute cannot be absolutely met, validation principles were inculcated into the software to ascertain as much as possible that data entering or leaving the software database was reliable. Data loss, which is the unforeseen loss of data or information in a computer system, can be very disastrous; this was also very much taken into account. Studies have consistently shown hardware failure and human error to be the most common causes of data loss. To avoid such disasters, a BackUp module was integrated. The database can be uploaded through a network stream onto an internet server for backup. Where there is no network available, an external device could be used for backing up data in a very fast and simple way.

*C. Integrating GIS*

The BeaconBase application connects to a GIS dataset to query geographic data. The algorithm used to match the Beacon feature in the GeoDatabase is depicted graphically in Fig. 2.

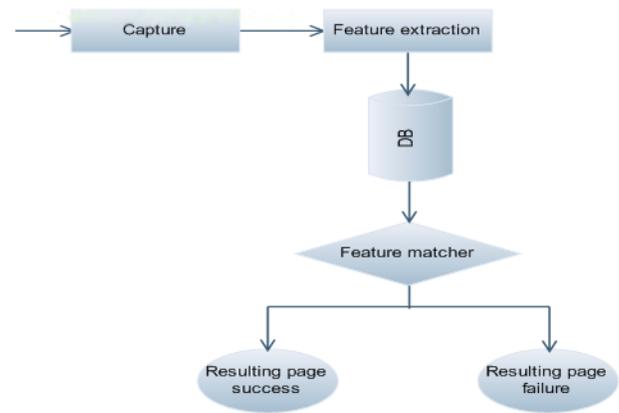

Fig. 2: Conceptual algorithm for connecting to GIS

To bind DataControl to data the Expression Blend software was used to automatically generate XAML code for that purpose. Classes describe the structure of objects and are a fundamental building block of object-oriented programming. Four object classes were created: winBackUp, winAbout, winBeaconBase and winSettings, all of which inherits from the base class System.Windows.Window.

### III. RESULTS AND ANALYSIS

*A. The InstruVisio Application*

The developed InstruVisio application has the following features: database of all instruments in the instrument store; a vivid description of all instruments and their use; visualization of all rooms through a video display; the ability to show the shelf location of each instrument in the store; a display of job types and instruments needed for each job; a built-in search engine to help determine with ease the location of instruments. An audio module has been incorporated using Microsoft speech synthesizer. It reads out, on request, the description of each instrument in the store. The application runs with a graphical user interface that is user-friendly and it afford the user total control. It displays an aesthetically pleasing view of buttons and an animated pop-up of windows to show the rooms in the Store, as depicted in Fig. 3.

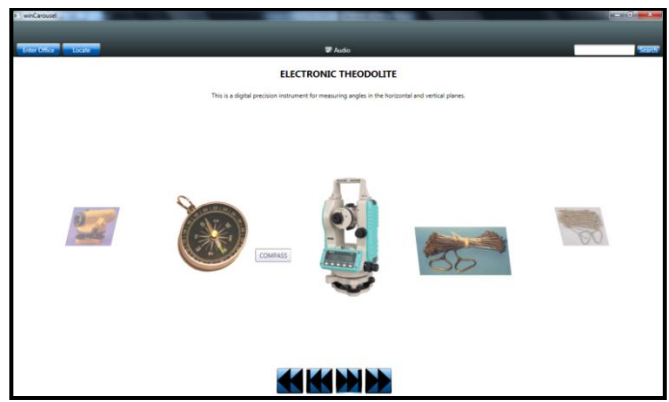

Fig. 3: The InstruVisio Application interface

The program starts with a video display to show the current state of the Equipment Store. Individual videos of each room can also be played on request by the user. The program is such that any new user can visualize the instrument store without necessarily going there. A Carousel displays in a window all instruments in the store and their description.

*B. The BeaconBase Application*

The BeaconBase application has the following features: database of all the survey Beacons in the study area; management features, such as adding a new beacon, editing a Beacon's information, Updating Beacon area photo; exporting beacon data locally onto a disk drive and printing in hardcopy; performing bearing and distance computation between any two beacons; visualizing beacon in a GIS window; viewing Beacon coordinates in various units; and a fast search engine algorithm. The application launches with a user-friendly interface. All the beacons in the study area are loaded from the database into memory. The customized data template of a listbox in the left panel is populated with a collection object. The user can select a Beacon from the list and instantly get information about the beacons - Coordinates, Picture and Revision Date. These data can be downloaded onto a local drive or printed out. The selected beacon can be visualized graphically in a GIS map. The user can add new beacons to the collection or edit data on the already existing beacons from the user-friendly interface. The spatial locations of the beacons can also be integrated with a GIS. Fig. 4 shows the BeaconBase Application interface.

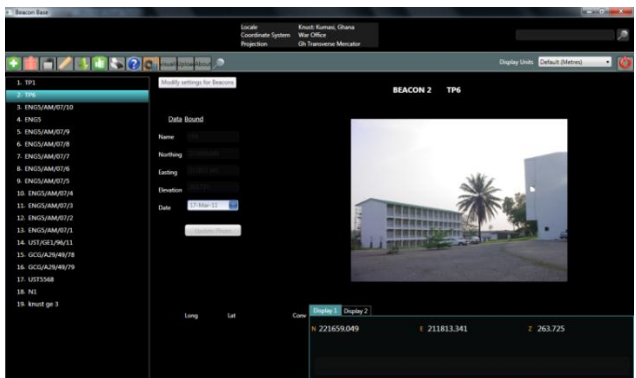

Fig. 4: The BeaconBase Application interface

*C. The LendingBase Application*

The LendingBase application has the following features: a form for recording new lending with details of borrower and all instruments borrowed; issuing a return note on a lending after the borrower has returned all instruments; viewing the lending history in a flexible and intuitive chart panel; editing instrument information in database; graphically generating the recent lending statistics - this statistics is very important because it highlights the rate of collection of each instrument in the Store; safe deletion of unwanted records - safe deletion here refers to the records not being totally cleared but transferred into a recycle bin so that it can be reviewed at a later time in case there is a problem. It is important to make it possible for deleted data to be recalled/restored to back-up database onto the web server or some other external storage device. This facilitates retrieval of data in case of damage to computer hardware. It is especially important because it helps prevent (or provide the necessary remedy for) program failure when it occurs due to data loss. An audio module has been incorporated to provide an audio rendition of the status of transactions. A built-in Search engine makes it possible to search for transactions that were executed in the past, and so help to study the trend of use and, therefore, the possible causes of what faults might result with any particular instrument.

After the application launches, all the lending records in the database are loaded into memory and populated into the DataGrid. The user can select a record from the DataGrid, which contains the name of the borrower, department, date of borrowing, phone number, return status and date of return. Details about the lending such as all instruments lent, quantity and corresponding serial numbers are displayed concurrently in the bottom panel. The application was designed with the aim of placing the user in absolute control. The user can add new records, with all the details pertaining to the record, in the AddNew Form. Thus, the user can see at a glance a pie chart showing in relative terms the quantity of instruments currently available in the Equipment Store, instruments lent out, and faulty instruments. The user can safely delete any record. Such deleted record will not be completely deleted from the database, but will be moved to a temporary space from where it may be restored whenever so desired. The borrower can be issued a return note on the lending transaction after he/she has returned all instruments. The quantity of available instruments in the database is automatically updated. The trend in recent lending activity in the Equipment Stores can be depicted graphically in a chart - a line chart, bar chart or pie chart. The data points of Fig. 5 shows a line chart of the total numbers of instruments lent out from the office in a day.

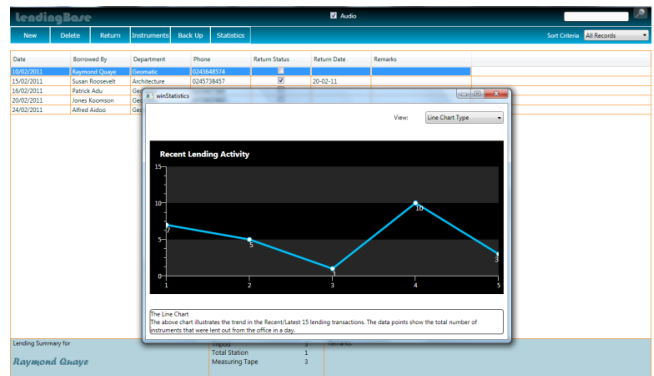

Fig. 5: The Lending Application interface

*D. The SurvCom Application*

The SurvCom application consists of modules for the following computations: area, detailing, horizontal curve setting out parameters derivations, and leveling - for both rise & fall method and height of collimation method. These computational types have been included because they are the most common day-to-day computational routines of the land surveyor. However, the system has the capability of adding other computational procedures. The Area Computation application calculates area by coordinates and, also, facilitates conversion from one unit of area to another. For example one can convert from square metres to hectares. Fig. 6 shows the Area Computation Application interface.

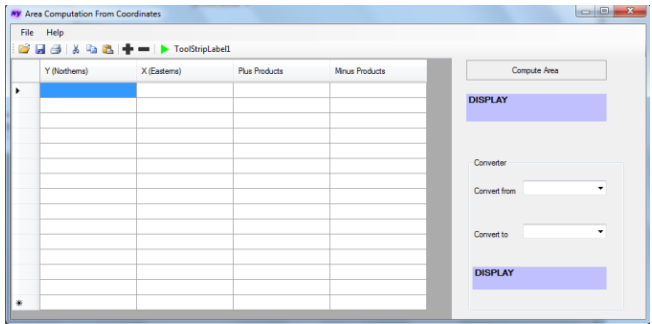

Fig. 6: The Area Computation Application interface

The Detailing Application uses the method of detailing by rays and can compute coordinates of practically infinite number of points and from an infinite number of instrument stations (Fig. 7). Use is made of observational values taken from field with a Total Station, which include the horizontal and vertical circle readings, and the slope distances. These are used to compute the coordinates of the points in question.

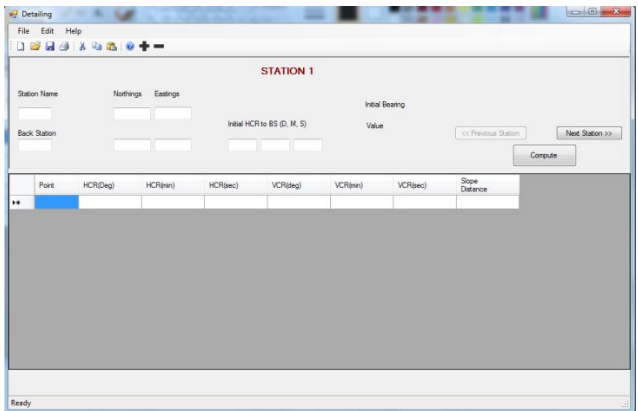

Fig. 7: The Detailing Application interface

The Horizontal Circular Curve Design Application computes the setting-out parameters of a horizontal circular curve. The required input values include deflection angle for the curve, curve length or radius of curvature, chainage of the intersection point for the forward and back tangents, as well as multiples of chord lengths for setting-out the curve (Fig. 8).

The Leveling Application computes the reduced level from back-sight, foresight and inter-sight values. It computes the reduced level using the rise & fall method or height of collimation method depending on the user's preference (Fig. 9). The developed Suite was subjected to considerable testing. The test participants included lecturers and students from the Department of Geomatic Engineering. The final design was based on feedback from test participants. The design considerations included a requirement for the suite to run optimally under Microsoft Windows Vista or higher Operating System; Graphics Application Programming Interface (API): DirectX; Random Access Memory (RAM) of at least 256MB; audio output devices like speakers. The compiled suite was built with MS Visual Studio 2008 Service Pack 1; Telerik Rad Controls for WPF version Q2 2009; ESRI's ArcObjects runtime license; and Developer Express v2009 vol1 controls. Running LendingBase initiates a window that pops up to display a Table with a design Form of the record keeping requirements of the Equipment Store.

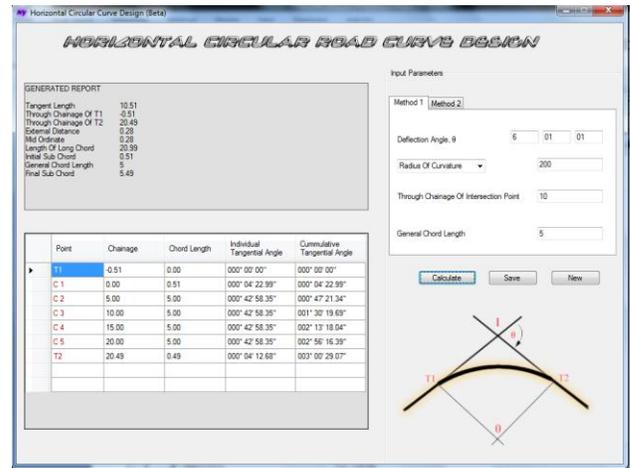

Fig. 8: The Horizontal Circular Curve Design Application interface

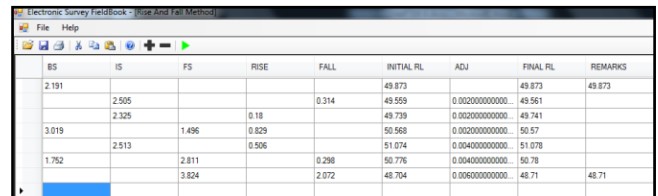

Fig. 9: The leveling Application interface

The menu bar has options for new entry and return of transaction, backing up data on a web server or on an external storage device, quantity of available instruments for lending, statistical data on recent issuing out of instruments and deletion of data in case of error in the entry. Summary of all details of each transaction is displayed on a record-details panel found at the bottom. These options allow the user of the program to be in control of every facet of instrument lending in the Equipment Store. Other Information about the program makeup can be found in the "About" section of the menu bar.

The BeaconBase program displays an artistic window with a well-organized menu bar. The menu bar buttons include Add, Delete, Save, Print, Edit, Visualize, Mark, Help, Back up and Export Data. It also has a drop-down button that displays units of beacons with which the user may want to work in. The Visualize button shows, in a GIS environment, the spatial location of the beacon. There is also a section with the list of beacons on KNUST campus; a single click on any beacon in the list displays the beacon's coordinates as well as a photograph of it to depict its current state. Any other information about the program is found in the About section of the menu bar. The InstruVisio Application launches with a window which plays a video coverage of the Equipment Store. There are buttons on the menu bar for the Carousel, EnterOffice, Help and About. The Carousel gives information and location of all instruments in the office. The EnterOffice is for visualizing any room in the office and shows the shelf locations of instruments. There is also a drop-down button which displays the various survey job types and the instruments required for such jobs. This is meant to assist the user in his/her choice of instruments for the survey work on hand.

This SurvCom program consists of a set of applications that automate certain survey computations. The Area

Computation is an application which with a click of a button, computes the area of a parcel from given coordinates. Its menu bar has Open Document, Add and Remove, Save, Cut, Paste and Compute buttons. The Detailing Application has a window with textboxes in which the initial Station name and coordinates with horizontal circle readings are input. In the Table, subsequent horizontal circle readings and vertical circle readings of targets will be entered as well as their slope distance. The Compute button in the program will then initiate the computations of all coordinates of points which can be plotted for detailing. The horizontal circular curve design application has a window with a section where the user may enter the deflection angle for a curve, curve length or radius of curvature, chainage of the intersection point for the forward and back tangents. The application then generates chord lengths for setting out the curve. The user may then enter the point names, their chord lengths, chainages and their individual tangential angles. Upon clicking the Calculate button, the application will automatically generate the cumulative tangential angles, give a sketch of the look of the curve and provide a general report on the work. The general report includes the through chainage of the start and end points, initial and final sub-chord lengths, the external distance, tangent length, and length of long chord. The Leveling Application displays a window with a menu bar. The program buttons are for Add, Remove, Open document, Cut, Paste, Help and Run. When File is clicked, options on method of computation pop up. These methods are the Rise & Fall method and the Height of collimation method. Depending on the user's discretion any method can be selected for computing the reduced levels.

## IV. CHALLENGES AND DISCUSSIONS

During the collection of data from the Equipment Office, considerable problems were encountered due to some of the analogue procedures employed at the time. For example, as regards the information on the beacons, there were problems with some of the coordinates since some characters were not legible on the printouts due to the age paper medium of storage. They were old, tearing and deteriorating. These realizations helped to appreciate the relevance of this work the more.

## V. CONCLUSION

A digital information and management system for the Equipment Store of the KNUST Department of Geomatic Engineering has been produced. That is, the objective of the research of creating a database of instruments in the Equipment Store with a GUI was achieved. This incorporated a fully scalable GUI database of beacons on KNUST campus integrated with its spatial location; a software to manage and archive the record keeping of instruments in a digital environment; and lastly automation few survey computations. The database for all the instruments in the office gives relevant descriptive information of each instrument in a very comprehensive and user-friendly GUI. A fully scalable GUI database of beacons on KNUST campus has also been developed and successfully integrated in a GIS geodatabase to provide spatial and graphical data for the beacons. The LendingBase Application manages, archives and records instruments and their daily lending transactions in a digital environment. The SurvCom provides a predefined automation of some survey computations for the user. The modules of the applications are user-friendly and provide an efficient and effective way of managing survey instruments and records to help sustain the activities and contributions of the Land Surveyor. Although the system in its current form can only be said to be prototypical, it can be modified or extended to cover any Equipment Office practice or extent of land coverage. Any Survey Firm, private or public, where records need to be kept on survey beacons and on the daily use of survey equipment can use this Suite to manage the relevant concerns. The importance of good record-keeping on surveyors' beacons and equipment cannot be over-emphasized as our dear nation enters the efficiency-demanding realm of oil and gas exploration. The BeaconBase Application can be employed by small or large organizations where coordinates data are collected, archived and used; such as with the use of the Global Positioning System (GPS). This can support decision makers in area where monitoring, such as grassland, wetlands, desertification, drought, etc. where beacon database are needed for decision makers in their investigations.

## REFERENCES

[1] Brase, J. and Farquhar, A., Access to Research Data, D-Lib Magazine, The Magazine of Digital Library Research, 17(1/2), 2011.
[2] Crosas, M., The Dataverse Network: An Open-Source Application for Sharing, Discovering and Preserving Data, D-Lib Magazine, The Magazine of Digital Library Research, 17(1/2), 2011,
[3] Galindo, J., (Ed.), Handbook on Fuzzy Information Processing in Databases, Hershey, PA: Information Science Reference, (an imprint of Idea Group Inc.), 2008.
[4] The Tech-FAQ , What is a Database? Electronic Archive: http://www.topbits.com/what-is-a-database.html, Accessed 24/07/2012, 2012
[5] Connolly, T. M. and Begg, C. E., Database Systems: A Practical Approach to Design, Implementation and Management, Addison Wesley Publishing Company, 2004
[6] Date, C. J., An Introduction to Database Systems, Addison Wesley Longman, 2003.
[7] Teorey, T., Lightstone, S. and Nadeau, T., Physical Database Design: the database professional's guide to exploiting indexes, views, storage, and more. Morgan Kaufmann Press, 2007
[8] Shih, J. Y., Why Synchronous Parallel Transaction Replication is Hard, But Inevitable? A White Paper, Parallel Computers Technology Inc. (PCTI), U.S.A. Electronic Archive: http://www.pcticorp.com/assets/docs/PQL2b.pdf. Accessed: 23/07/12, 2007
[9] Thomas, D., Hunt, A., and Thomas, D., Programming Ruby: A Pragmatic Programmer's Guide, Addison-Wesley Professional; 1st edition, 2000
[10] Foster, C., Pennington, C. V. L., Culshaw, M. G., and Lawrie, K., The national landslide database of Great Britain: development, evolution and applications, Environmental Earth Sciences, 66(3), pp. 941-953
[11] Dahl, R., Wolden, K., Erichsen, E., Ulvik, A., Neeb, P. R. and Riiber, K., Sustainable management of aggregate resources in Norway, Bulletin of Engineering Geology and the Environment, 71(2), pp. 251-255
[12] Steele, C. M., Bestelmeyer, B. T., Burkett, L. M., Smith, P. L., and Yanoff, S., Spatially Explicit Representation of State-and-Transition Models, Rangeland Ecology & Management, 65(3), pp. 213-222
[13] Panigrahy, S., Ray, S. S., Manjunath, K. R., Pandey, P. S., Sharma, S. K., Sood, A., Yadav, M., Gupta, P. C., Kundu, N., and Parihar, J. S., A Spatial Database of Cropping System and its Characteristics to Aid Climate Change Impact Assessment Studies, Journal of the Indian Society of Remote Sensing, 39(3), pp. 355-364
[14] Fu, P. and Sun J. (2010). Web GIS: Principles and Applications, ESRI Press, Redlands, CA